# Persian phonemes recognition using PPNet

Saber Malekzadeh[a], Mohammad Hossein Gholizadeh[*], Seyed Naser Razavi[b]

a. Department of Computer Sciences, Faculty of Mathematics and Computer Sciences
Vali-e-Asr University of Rafsanjan
b. Department of Computer Engineering, Faculty of Computer & Electrical Engineering
University of Tabriz

*Abstract*— **In this paper a new approach for recognition of Persian phonemes on the PCVC speech dataset is proposed. Nowadays deep neural networks are playing main rule in classification tasks. However the best results in speech recognition are not as good as human recognition rate yet. Deep learning techniques are shown their outstanding performance over so many classification tasks like image classification, document classification, etc. Also in some tasks their performance were even better than human. So the reason why ASR (automatic speech recognition) systems are not as good as the human speech recognition system is mostly depend on features of data is fed to deep neural networks.
In this research first sound samples are cut for exact extraction of phoneme sounds in 50ms samples. Then phonemes are grouped in 30 groups; Containing 23 consonants, 6 vowels and a silence phoneme. STFT (Short time Fourier transform) is applied on them and Then STFT results are given to PPNet (A new deep convolutional neural network architecture) classifier and a total average of 75.87% accuracy is reached which is the best result ever compared to other algorithms on Separated Persian phonemes (Like in PCVC speech dataset).**

1. INTRODUCTION

In the 4th industrial revolution, smart machines have become a part of our modern life and this has encouraged the expectations of friendly interaction with them. The speech, as a communication way, has seen the successful development of quite a number of applications using automatic speech recognition (ASR), including command and control, dictation, dialog systems for people with impairments, translation and etc. But, the actual challenge is to create inputs and use of speech to control applications and access information. Research on ASR is still an active topic to use speech as an input [1].

ASR – the recognition of the information of a speech signal and its transcription to a set of characters – is the object of research for more than five decades, achieving notable results. It is only to be expected that advances in speech recognition make speech in any language, as the best input method when the recognizers reach error rates under 5%. While digit recognition has already reached a rate of 99.6% [2], the phoneme recognition has not gone far more than 83% [3].

In any large vocabulary LVASR (Large vocabulary automatic speech recognition) systems, more than language model, the performance, depends on phoneme recognizer. This is why research groups still working on developing the better phoneme recognizer systems. The phoneme recognition is in fact, a recurrent problem for speech recognition community.

Phoneme recognition can be found in many applications nowadays. In addition to some typical LVASR systems [4], it can be found in applications related to language and speaker recognition, music identification and also translation.

The challenge of building good acoustic models starts with applying good training algorithms to a suitable set of data. The dataset contains sound units which can be trained on training algorithms and is dependent to the detail of annotation of those units. Most of datasets are not labeled at the exact phoneme level [5]. Thus, the PCVC dataset is employed because it is labeled at phoneme level. Also, unlike other phoneme based datasets [6], PCVC contains just 2 phoneme in every sample which makes the training and recognition process better. For extraction of consonants, as they are just pronounced before vowels, it is possible to separate them approximately. However, in this paper, to show the usability of phoneme recognition on PCVC, the recognition algorithms are examined on PCVC to get the best recognition results.

This article is organized as follows. Section 2 discusses about the PCVC speech dataset which is used and presented in phoneme recognition task for the first time in this paper. Section 3 proposes the preprocessing level with sound signal processing algorithms like STFT and phoneme extraction from PCVC speech dataset. Section 4 describes the deep artificial neural network which is applied for samples classification. Section 5 brings the conclusion and the last section is acknowledgement.

2. PCVC SPEECH DATASET

This dataset contains of 22 Persian consonants and 6 vowels which is listed in table 1. Table 1 contains the vowels and consonants just like the dataset. There are 12 speakers including 5 males and 5 females and 1 child. All sound samples are possible combinations of vowels and consonants (132 samples for each speaker). The sample rate of all 2 seconds speech samples is 48000 which means there are 48000 audio samples (values) in every second. Every sound sample is 2 seconds speech sample, which in average, 0.5 second of each sample is speech and the rest is silence.

For testing process, 15 percent of samples are selected, and 85 percent of those are used in training process. [7]

*Table 1- Phoneme List in PCVC dataset*

| Persian form | English form | Persian Example |
|---|---|---|
| آ | A | آل |
| ای | I | ایل |
| او | ʊ | او |
| اَ | æ | اول |
| اِ | e | اسم |
| اُ | o | اردو |
| پ | P | پا |
| ب | B | با |
| ت | T | تا |
| د | D | دارو |
| چ | tʃ | چاقو |
| ج | dʒ | جارو |
| ک | K | کاری |
| گ | G | گاری |
| ف | F | فاطمه |
| و | V | واهمه |
| خ | Kh | خاطره |
| س | S | ساز |
| ز | Z | زار |
| ش | ʃ | شار |
| ژ | ʒ | ژاکت |
| م | M | ماکت |
| ن | N | نادی |
| ه | H | هادی |
| ل | L | لابه |
| ر | R | راهبه |
| ق | Q | قاری |
| ی | j | یاری |

## 3. SPEECH SIGNAL PREPROCESSING

### A. Phoneme extraction

Every sound sample is an audio wave of 2 seconds of speech which ends with silence for at least 0.25s. Then, consonants and vowels are pronounced consecutive. Intensity of silence is almost (not exactly) zero. This values (Intensity of silence and consonants) can be employed to detect the vowels which have higher intensity than silence. Thus, the vowels are parts of the speech which their intensity is more than 0.25 of the max intensity of sound sample. 0.25 is a suitable benchmark for detection of vowels from other elements on PCVC dataset. This part of speech is enough to detect vowels in sound sample.

Each vowel sample is cut in 50ms samples. In this paper the aim was to recognize phonemes; therefor as said before vowels are pronounced just after consonants. So almost 50ms of speech before vowels is selected as consonant speech sample. Actually this splitting method is used on PCVC speech dataset. But in real world there are sentences which there is no rule for how phonemes are gathered together in them.

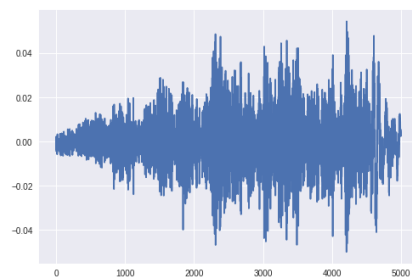

*Figure 1- A two phonemes sample time domain plot*

### B. Adaptive Noise Cancellation

An Adaptive Noise Canceller (ANC) has two inputs – primary and reference. The primary input receives a signal s from the signal source that is corrupted by the presence of noise n uncorrelated with the signal. The reference input receives a noise $n_0$ uncorrelated with the signal but correlated in some way with the noise n. The noise $n_0$ passes through a filter to produce an output nˆ that is a close estimate of primary input noise. This noise estimate is subtracted from the corrupted signal to produce an estimate of the signal at sˆ, the ANC system output. [8]

### C. Short time Fourier transform

In this paper, Continues STFT algorithm is used to extraction of features from sound samples. STFT is used as one of the best sound feature extraction algorithms for decades.

STFT is a sound feature extraction algorithm which gives the ability of transform sound features from temporal domain to frequency domain and from frequency domain to time-frequency domain [9]. On time-frequency diagram, some features like shape of formants, distance, sound shocks and also Curvature of formants can be found that are the vital features for phoneme recognition. The mentioned features are the results of human local folds, lips, tongue and teeth which are always creating patterns in time-frequency domain in different shapes when phonemes are being pronounced [10] [11].

STFTs are commonly derived as follows:

1. Separating sound samples in fixed size intervals.
2. Applying Fourier transform on each sound interval.
3. Assortment of frequencies in different frequency ranges.
4. Initializing of each time-frequency domain point (rectangle) with suitable values based on their number of samples. [12]

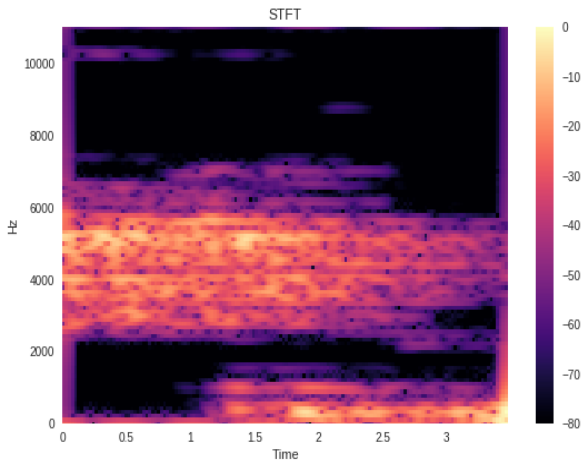

*Figure 2- A phoneme sample time-frequency domain plot*

In this paper, this algorithm is utilized for extraction of the spectral features of sound samples. For this purpose, the window length of 5ms. Each window includes 150 frequency ranges. These parameters are identified as a suitable choice in tests. The reason why STFT is chosen is because of experience. MFCC, STFT and raw sound sample are tested in classification level on some of samples and the best results were achieved in STFT. It can be argued that because STFT is simple FFT on sound intervals, in non-noise environments STFT is better resulted than others. STFT Mathematically, is written as:

**STFT**$\{x(t)\}(\tau, \omega)$ dt

Where w(t) is the window function, centered around by zero, and x(t) is the signal to be transformed. $X(\tau, \omega)$ is essentially the Fourier Transform of $x(t)w(t-\tau)$, a complex function representing the phase and magnitude of the signal over time and frequency. The time axis is $\tau$ and the frequency axis is $\omega$. [13]

### 4. DEEP NEURAL NETWORK

A deep neural network is a kind of artificial neural network which has many layers to perform better classification (usually artificial neural networks with more than 3 layers are called deep networks.). One kind of layers in deep neural network structures is convolutional layer and network with convolutional layers commonly can be named as convolutional neural network which is used by so many researchers in different tasks like classification problems. In a convolutional neural classifier there are some inputs and desired outputs to train and test classes [14]. Each convolutional neural network has also other kind of layers like Pooling layers, Activation layers, Fully-connected layers and etc.

*A. Training process*

To train phoneme speech samples, PPNet (A new architecture as a convolutional artificial neural network) is used. The architecture used is provided in figure 4.

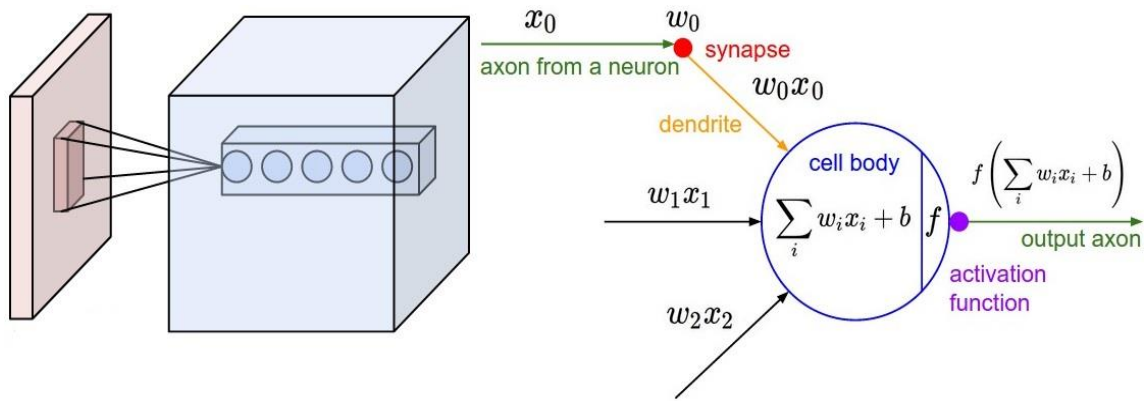

*Figure 3- Convolution Structure [15]*

```
Layer (type)                 Output Shape              Param #
=================================================================
conv2d_1 (Conv2D)            (None, 100, 150, 32)      320
batch_normalization_1 (Batch (None, 100, 150, 32)      128
activation_1 (Activation)    (None, 100, 150, 32)      0
dropout_1 (Dropout)          (None, 100, 150, 32)      0
conv2d_2 (Conv2D)            (None, 100, 150, 32)      9248
max_pooling2d_1 (MaxPooling2 (None, 50, 75, 32)        0
conv2d_3 (Conv2D)            (None, 50, 75, 64)        18496
dropout_2 (Dropout)          (None, 50, 75, 64)        0
conv2d_4 (Conv2D)            (None, 50, 75, 64)        36928
max_pooling2d_2 (MaxPooling2 (None, 25, 37, 64)        0
conv2d_5 (Conv2D)            (None, 25, 37, 128)       73856
dropout_3 (Dropout)          (None, 25, 37, 128)       0
conv2d_6 (Conv2D)            (None, 25, 37, 128)       147584
max_pooling2d_3 (MaxPooling2 (None, 12, 18, 128)       0
flatten_1 (Flatten)          (None, 27648)             0
dropout_4 (Dropout)          (None, 27648)             0
dense_1 (Dense)              (None, 1024)              28312576
dropout_5 (Dropout)          (None, 1024)              0
dense_2 (Dense)              (None, 128)               131200
dropout_6 (Dropout)          (None, 128)               0
dense_3 (Dense)              (None, 30)                3870
=================================================================
```

*Figure 4- CNN model summary*

6 convolutional layers are used all with stride (1, 1), kernel size (3, 3) and Relu activation function. First two convolution layers have 32 kernels, second two have 64 and third two have 128 kernels.

Convolution layer instructions are as follows:
- Accepts a volume of size $W1 \times H1 \times D1 W1 \times H1 \times D1$
- Requires four hyper parameters:
    - Number of filters KK,
    - Their spatial extent FF,
    - The stride SS,
    - The amount of zero padding PP.
- Produces a volume of size $W2 \times H2 \times D2 W2 \times H2 \times D2$ where:
    - $W2=(W1−F+2P)/S+1 W2=(W1−F+2P)/S+1$
    - $H2=(H1−F+2P)/S+1 H2=(H1−F+2P)/S+1$ (i.e. width and height are computed equally by symmetry)
    - $D2=K D2=K$
- With parameter sharing, it introduces $F \cdot F \cdot D1 F \cdot F \cdot D1$ weights per filter, for a total of $(F \cdot F \cdot D1) \cdot K (F \cdot F \cdot D1) \cdot K$ weights and KK biases.
- In the output volume, the dd-th depth slice (of size $W2 \times H2 W2 \times H2$) is the result of performing a valid convolution of the dd-th filter over the input volume with a stride of SS, and then offset by dd-th bias. [15]

A batch normalization layer is used after first convolutional layer. Also 6 dropout layers are used to avoid overfitting. Also three maxpooling layers are used to help the network learn data orientation and also general features. Batch size was 16 and number of epochs was 50. Input shape was 100*150. [16]

*B. Testing process*

Test data was selected fully randomly from training data with a guaranty of selecting between 8 and 16 from each class of data for test set. Results for best selected phonemes are shown in Table 2.

**Precision = True_Positive / (True_Positive + False_Positive)**

**Recall = True_Positive / (True_Positive + False_Negative)**

**F1_score= (2* Precision* Recall) / (Precision + Recall)** [17]

The overall F1_score result was 75.78% of recognition accuracy for all phonemes. In Table 2, first 23 phonemes are consonants with order of table 1, the rest 6 phonemes are vowels again like the order in table 1 and the last phoneme is silence.

## 5. CONCLUSION

In this paper a new method is proposed for recognition of phonemes in Persian language on PCVC. This method can be used not only for recognizing of mono-phonemes but also can be used as an input to selecting best words in speech transcription. As results shown in Table 2, the capability of the proposed vowels and consonants recognition system in predicting the phonemes are better than the other phoneme recognition methods which are proposed in the conventional approaches.

*Table 2- Results Table*

|    | Precision | Recall | F1_score | Support |
|----|-----------|--------|----------|---------|
| 0  | 0.78      | 0.58   | 0.67     | 12      |
| 1  | 0.70      | 0.58   | 0.64     | 12      |
| 2  | 0.78      | 0.64   | 0.70     | 11      |
| 3  | 0.59      | 0.83   | 0.69     | 12      |
| 4  | 0.54      | 0.64   | 0.58     | 11      |
| 5  | 0.58      | 0.92   | 0.71     | 12      |
| 6  | 0.75      | 0.69   | 0.72     | 13      |
| 7  | 0.80      | 0.73   | 0.76     | 11      |
| 8  | 0.91      | 0.67   | 0.77     | 15      |
| 9  | 0.67      | 0.50   | 0.57     | 16      |
| 10 | 0.62      | 0.80   | 0.70     | 10      |
| 11 | 0.69      | 0.69   | 0.69     | 13      |
| 12 | 0.79      | 0.69   | 0.73     | 16      |
| 13 | 0.58 | 0.78 | 0.67 | 9 |
| 14 | 0.74 | 0.88 | 0.80 | 16 |
| 15 | 0.60 | 0.50 | 0.55 | 12 |
| 16 | 0.62 | 0.56 | 0.59 | 9 |
| 17 | 0.77 | 0.77 | 0.77 | 13 |
| 18 | 0.54 | 0.50 | 0.52 | 14 |
| 19 | 0.75 | 0.67 | 0.71 | 9 |
| 20 | 0.75 | 0.90 | 0.82 | 10 |
| 21 | 0.73 | 0.67 | 0.70 | 12 |
| 22 | 0.78 | 0.78 | 0.78 | 9 |
| 23 | 0.90 | 1.00 | 0.95 | 9 |
| 24 | 1.00 | 1.00 | 1.00 | 14 |
| 25 | 1.00 | 1.00 | 1.00 | 13 |
| 26 | 1.00 | 0.94 | 0.97 | 16 |
| 27 | 1.00 | 1.00 | 1.00 | 9 |
| 28 | 1.00 | 1.00 | 1.00 | 13 |
| 29 | 1.00 | 1.00 | 1.00 | 9 |

6. ACKNOWLEDGMENT

So many thanks to those helped us to develop PCVC dataset especially speakers: Farideh Jabraili, Hedayat Malekzadeh, Hamed Afjuland, Mohammad Ataeizadeh, Tahereh Salari, Alireza Aghaei, Parisa Seyfpour, Sahel Soltani and Mina Bayarash.

Also especial thanks to Prof.Beigi for their good information that helped us in this project.